\documentclass[a4paper,11pt]{article}
\pdfoutput=1
\usepackage{amsmath}
\usepackage{amssymb}
\usepackage{color}
\usepackage{cite}
\usepackage{hyperref}
\usepackage{graphicx}
 \usepackage{leftidx}
 \usepackage{subfig}

\makeatletter
\@addtoreset{equation}{section}
\renewcommand{\theequation}{\thesection.\@arabic\c@equation}
\makeatother
\makeatletter
\renewcommand\appendix{\par%\newpage
  \setcounter{section}{0}%
  \setcounter{subsection}{0}%
  \gdef\thesection{Appendix \@Alph\c@section }
  \renewcommand{\theequation}
  {\Alph{section}.\arabic{equation}}
}
\makeatother
\def \be {\begin{equation}}
\def \ee {\end{equation}}
\def \ba {\begin{array}}
\def \ea {\end{array}}
\def \bea{\begin{eqnarray}}
\def \eea{\end{eqnarray}}

\def \a {\alpha}

\def \g {\gamma}

\def \m {\mu}
\def \n {\nu}

\def \nn {\nonumber}

\def \hs {\hspace}

\def \inf {\infty}

\def \Tr {{\textrm{Tr}}}

\def \cW{\cal W}

\def \tr {{\textrm{tr}}}

\title{\textbf{Holographic Entanglement Entropy For a Large Class of States in 2D CFT}}
\author{
Bin Chen$^{1,2,3}$\footnote{bchen01@pku.edu.cn}\,
and
Jie-qiang Wu$^{1}$\footnote{jieqiangwu@pku.edu.cn}
}
\date{}

\begin{document}

\maketitle

\begin{center}
{\it
$^{1}$Department of Physics and State Key Laboratory of Nuclear Physics and Technology, Peking University, Beijing 100871, P.R.\! China\\
\vspace{2mm}
$^{2}$Collaborative Innovation Center of Quantum Matter, 5 Yiheyuan Rd, \\Beijing 100871, P.~R.~China\\
$^{3}$Center for High Energy Physics, Peking University, 5 Yiheyuan Rd, \\Beijing 100871, P.~R.~China
}
\vspace{10mm}
\end{center}

\begin{abstract}

In this paper, we study the  entanglement entropy in a large class of states of two-dimensional conformal field theory in the the large central charge limit.  This class of states includes the states created by the insertion of a finite number of local heavy operators. By using the monodromy analysis, we obtain the leading order entanglement entropy for the general state. We show that it is exactly captured by the Ryu-Takayanagi formula, by using the Wilsonian line prescription in the Chern-Simons formulation of the AdS$_3$ gravity. %The discussion can be extended to the higher spin theory. In higher spin theory, we study two special descendant operators correlation function and we show that this two point function can be described by a Wilson line prescription in five dimension.

\end{abstract}

\baselineskip 18pt
\thispagestyle{empty}

\newpage

\section{Introduction}

The entanglement entropy  captures the quantum entanglement between a system and its environment\cite{nielsen2010quantum,petz2008quantum}. In a field theory which has infinite degrees of freedom, the entanglement entropy is quite hard to evaluate. Based on the AdS/CFT correspondence, it has been suggested that for the conformal field theory(CFT) with a gravity dual, the entanglement entropy can be evaluated holographically by the Ryu-Takayanagi(RT) formula\cite{Ryu:2006bv,Ryu:2006ef} in the semi-classical limit
\be S_{HEE}=\frac{A}{4G}.\ee
$A$ is the area of the minimal surface in the bulk which is homologous to the entangling region. The holographic entanglement entropy implies a deep relation between the quantum entanglement and the gravity. It opens a new window to study the AdS/CFT correspondence. The RT formula is reminiscent of the Bekenstein-Hawking formula for the black hole entropy. Actually in its first derivation for a spherical region in the vacuum state of a CFT, it could be mapped to the computation of the entropy of a hyperbolic black hole in the bulk\cite{Casini:2011kv}. Furthermore, by using the replica trick in the gravity side and the Euclidean gravity action, the RT formula was proven for general regions following the the AdS/CFT correspondence. The holographic entanglement entropy could be taken as  the generalized gravitational entropy \cite{Lewkowycz:2013nqa}.

In particular, in the context of AdS$_3$/CFT$_2$ correspondence, the holographic computation of the entanglement entropy is exact in the following sense. In the correspondence, the central charge of the dual CFT is $c=\frac{3l}{2G}$ so that the semiclassical gravity corresponds to the large $c$ limit of the dual CFT. Consequently the leading order entanglement entropy is captured by the RT formula \cite{Headrick:2010zt,Faulkner:2013yia,Hartman:2013mia} and the next-to-leading one is captured by the 1-loop correction to the RT formula\cite{Faulkner:2013ana,Barrella:2013wja}.  This picture  could be proved for the vacuum state of the CFT under reasonable assumptions.

On the bulk side, the on-shell regularized Euclidean action is reduced to the Liouville type action at the asymptotic boundary\cite{Krasnov:2000zq}. The classical boundary action could be determined by solving the monodromy problem of a differential equation\cite{Faulkner:2013yia,Hartman:2013mia}. At the 1-loop level, the picture relies on the fact that the 1-loop partition function of any handle-body configuration \cite{Yin:2007gv,Giombi:2008vd} could be reproduced by the CFT partition function\cite{Chen:2015uga}.

On the CFT side, by the replica trick\cite{Holzhey:1994we}, the computation of the entanglement entropy is  transformed into the partition function on a $n$-sheeted space pasted along the entangling surface. In $1+1$ dimension, the field theory on the $n$-sheeted surface can be regarded as a $n$-copied field theory on one surface with some twist operators being inserted at the branch points. For single interval in the full complex plane or the cylinder, by the conformal symmetry, the entropy is of an universal form, which is reproduced exactly by the holographic calculation\cite{Ryu:2006ef}. For more complicated entangling regions, the entanglement entropy  in general depends on the details of the theory, the spectrum and the OPE coefficients. However, for the CFT with a gravity dual, it turns out that the vacuum conformal block dominates the partition function. In the vacuum module, the states are constructed by acting the Virasoro generators on the vacuum. These states  in the large central charge limit correspond to the free gravitons in the bulk.

The techniques to compute the entanglement entropy in CFT$_2$ was formulated in \cite{Hartman:2013mia} following the monodromy prescription introduced in \cite{Zamolodchikov1,Zamolodchikov2}. In \cite{Zamolodchikov1,Zamolodchikov2} it was suggested that the semi-classical conformal block can be determined by solving the monodromy problem for a differential equation. For the two-heavy-two-light operator case the semi-classical conformal block can be solved explicitly \cite{Fitzpatrick:2014vua,Asplund:2014coa}. By taking a Wick rotation, this analysis can be used to study time-dependent effects including the thermalization and the scrambling effect\cite{Nozaki:2014hna,Caputa:2014vaa,Caputa:2014eta,Caputa:2015waa,Roberts:2014ifa,Fitzpatrick:2016thx}. The monodromy prescription can be generalized to more complicated cases, including the entanglement entropy on the torus \cite{Barrella:2013wja,Chen:2016uvu}, the one in the higher spin theory \cite{deBoer:2014sna,Chen:2016uvu} and the black hole formation\cite{Anous:2016kss}.

The exactness of the RT formula in AdS$_3$/CFT$_2$ has been proved or supported in a few cases. In the vacuum state of the CFT, the single interval case\cite{Ryu:2006ef} is trivial and  the multi-interval case has been proved in
\cite{Faulkner:2013yia,Hartman:2013mia}. For the single-interval on the torus case, though there are pieces of strong evidence to support the picture\cite{Chen:2014unl,Chen:2015kua,Chen:2015uia}, there is short of rigorous proof on the holographic computation, especially the size dependence suggested in \cite{Chen:2014unl}.  Furthermore the field theory computation  confirms the phase transition  suggested by the holographic study \cite{Ryu:2006ef,Azeyanagi:2007bj,Chen:2015kua}. It would be interesting to investigate the correctness of the RT formula in more general case.

In this work, we study the single-interval entanglement entropy in a large class of states of the large $c$ CFT. The  states  can be the vacuum state, the thermal state described by a thermal density matrix, the local quench state or any time-dependent or time-independent states generated by a finite number of local heavy operators\footnote{For the study of the excitation entanglement entropy, see \cite{Sheikh-Jabbari:2016znt}. }. By the replica trick, the computation of the entropy is transformed into the correlation function of two light operators  in the generated state. From the AdS$_3$/CFT$_2$ correspondence,  such kind of state in CFT is dual to a classical gravitational configuration, described by the Banados geometries.  We prove that the two-point correlation function can always be captured by the length of the geodesic  in the corresponding bulk background. Our proof relies on the Wilson-line prescription developed in \cite{Ammon:2013hba} to compute the holographic entanglement entropy.

%By using the monodromy prescription, the two point function can be directly related to Wilson line in 3d Chern-Simon form \cite{Ammon:2013hba}. Based on \cite{Ammon:2013hba}, when the gauge group is $SL(2)\times SL(2)$ the Wilson description equals to the geodesic length in gravity.

The remaining of the paper is organized as follows. In Section 2, we first briefly introduce the entanglement entropy and its computation in quantum field theory via the replica trick.  Then we  apply the monodromy techniques to compute the two-point function in a large class of states and derive  the monodromy equation, which is essential to read the two-point function. In section 3,  we  review the Wilson-line prescription to compute the HEE and  emphasize that the prescription leads to the RT formula in any bulk background.  Then we show that the on-shell action of the Wilson-line probe satisfy the monodromy condition, and thus gives the entanglement entropy correctly.  In section 4, we end with a conclusion and some discussions.

\section{Entanglement entropy in a large class of  states}

The entanglement entropy is a measure of the entanglement between the subsystem and its complement. Assuming the whole system is described by the density matrix $\rho$, we can define the reduced density matrix for the subsystem $A$ as
\be \rho_A=\tr_{A^{c}} \rho, \ee
where $A^{c}$ is the complement of $A$. The entanglement entropy of $A$ can be defined to be the von Neumann entropy
\be S_{EE}=-\tr \rho_A \log \rho_A. \ee
In the following discussion, we focus on the entanglement entropy in 2D CFT. By the replica trick, we can define the R\'enyi entropy
\be S_n=-\frac{1}{n-1}\log \frac{Z_n}{Z_1^n}, \ee
where $Z_n$ is the partition function on the $n$-sheeted Riemann surface obtained by pasting $n$-copies of the original system along the region $A$. If the state is excited by  some local operators, there should also be some operators inserted on the $n$-sheeted Riemann surface. Assuming the R\'enyi entropy can be analytically extended to non-integer value, the entanglement entropy can be read from taking $n\to 1$ limit,
\be S_{EE}=\lim_{n\rightarrow 1}S_n. \ee

In the path integral formalism, the partition function $Z_n$ can be understood in another way. Due to the replica symmetry, the path-integral over the field theory on the $n$-sheeted surface can be regarded as the path-integral of a $n$-copied field theory on one-sheet surface with the twist boundary condition on the fields at the branch points. The twist boundary condition can be induced by the twist operators inserting at the branch points in a orbifold CFT. Therefore there is
\be \frac{Z_n}{Z_1^n}=\langle {\cal{T}}(z_1){\cal{T}}(z_2)...\rangle. \ee
In the ellipsis $...$, there can be other twist operators or other local operators in a general state. The twist operator has a conformal dimension
\be h=\frac{c}{24}(n-\frac{1}{n}). \ee
When we take the $n\rightarrow 1$ limit to compute the entanglement entropy, the twist operator become a light operator. Therefore the entanglement entropy could be evaluated as the multi-point correlation function of the light operators .

For simplicity, let us consider the single interval case. In this case we just need to consider the correlation function of  two light operators. Taking the fusion of these two light operators, the correlation function can be expand into a series of the contributions from different conformal blocks. We assume that in the semi-classical limit,  the contribution from the vacuum conformal block dominates. %For multi-interval limit, there are no difficulties to extend our discussion for that case.

More generically, we may study the  correlation function on a Riemann surface
\be \langle \phi(z_1)\phi(z_2)\rangle_b\equiv\frac{\langle \phi(z_1)\phi(z_2)...\rangle\mid_{{\cal{R}}}}
{\langle ...\rangle\mid_{{\cal{R}}}} \label{light}\ee
where the operator $\phi$ is a light operator and the ellipsis $...$ include finite number of locally inserted heavy operators, and ${\cal{R}}$ can be any Riemann surface\footnote{For convenience, in the following discussion we just ignore the subscript ${\cal{R}}$. One should be aware that  the expectation value can be defined on any Riemann surface.}.  The conformal dimension of the operator $\phi$ satisfies $1<<h<<c$. Holographically this condition means that we can make the saddle point approximation and ignore the back-reaction. We can define the expectation value of the stress tensor  to be
\be\label{T0}  T_0(z)=\frac{\langle \hat{T}(z)...\rangle}
{\langle ...\rangle}. \ee
%In AdS/CFT correspondence, with the stress tensor expectation value, we can build the gravity solution close to the boundary. We will see the relation between field theory and gravity side.
In other words, the other operators in the ellipsis $...$ generate a general state, whose stress tensor is given by $T_0(z)$. Note that the stress tensor $T_0(z)$ is generically not a constant. The constant $T$ cases have been discussed in the literature \cite{Ryu:2006bv}.

The correlation function (\ref{light}) can be evaluated by using the
 the monodromy analysis \cite{Hartman:2013mia}.
Introduce a degenerate representation with null state
\be\label{null} \mid \chi\rangle =(L_{-2}-\frac{3}{2(2h+1)}L_{-1}^2) \mid \hat{\psi} \rangle, \ee
where
\be h=-\frac{1}{2}-\frac{9}{2c}, \ee
in the large $c$ limit. We define
\bea\label{2nd} &~&\psi(z)=\frac{\langle \hat{\psi}(z) \phi(z_1)\phi(z_2)...\rangle}
{\langle \phi(z_1)\phi(z_2)...\rangle }, \notag \\
&~& T(z)=\frac{\langle \hat{T}(z) \phi(z_1) \phi(z_2)...\rangle}
{\langle \phi(z_1)\phi(z_2)...\rangle}. \eea
Inserting the null state (\ref{null}) into the correlator
\be \langle \chi(z) \phi_1(z_1) \phi_2(z_2)...\rangle =0, \ee
we get the differential equation
\be \frac{\partial^2}{\partial z^2}\psi(z)+\frac{6}{c}T(z)\psi(z)=0 .\ee
This second order differential equation can be rewritten into a set of first order  differential equation
\be\label{equ1} \partial \Psi(z)=-a(z) \Psi(z), \ee
where
\be
\Psi(z)=\left(
\begin{array}{ccc}
-\frac{\partial}{\partial z}\psi_1(z)&-\frac{\partial}{\partial z}\psi_2(z) \\
\psi_1(z)& \psi_2(z)
\end{array}
\right),
\ee
\be
a(z)=\left(
\begin{array}{ccc}
0&-\frac{6}{c}T(z) \\
1&0
\end{array}
\right).
\ee 
$\psi_1$ $\psi_2$ are two independent solution for \ref{2nd}. 

By the Ward identity, the asymptotic behavior of the stress tensor near $z_1$ and $z_2$ is like
\bea &~&T(z)\sim \frac{h}{(z-z_1)^2} +\frac{\gamma_1}{z-z_1}+..., \notag \\
&~& T(z)\sim \frac{h}{(z-z_2)^2}+\frac{\gamma_2}{z-z_2}+.... \eea
We need to tune the parameter $\gamma_1$ and $\gamma_2$ such that the $\Psi(z)$ has proper monodromy around the cycle enclosing $z_1$ and $z_2$. Because we only choose the vacuum module  for the fusion of $\phi_1$ and $\phi_2$, the monodromy around the cycle enclosing $z_1$ $z_2$ should be trivial.

The equation (\ref{equ1}) can be  solved perturbatively. There is a solution $v_0(z)$ at the $0$-th order satisfying the equation
\be \frac{\partial}{\partial z}v_0(z)=-a_0(z)v_0(z), \label{v0}\ee
where
\be a_0(z)=\left(
\begin{array}{ccc}
0 & -\frac{6}{c}T_0(z) \\
1 & 0
\end{array} \right),
\ee
and $T_0(z)$ is the expectation value of the stress tensor  without inserting $\phi_1$ $\phi_2$, defined in (\ref{T0}).
The whole wavefunction is
\be
\Psi(z)=v_0(z)v_1(z)
\ee
and the function $a(z)$  can be expanded around the background $a_0(z)$
\be
a(z)=a_0(z)+a_1(z),
\ee
where
\bea &~&a_1(z)=a(z)-a_0(z)=\left(
\begin{array}{ccc}
0 & -\frac{6}{c}T_1(z) \\
0 &0
\end{array} \right), \eea
with
\be T_1(z)=T(z)-T_0(z). \ee
Because that $T_0(z)$ is not singular at $z_1$ and $z_2$, $T_1(z)$ has the same asymptotic behavior near $z_1$ and $z_2$ as $T(z)$
\bea &~&T_1(z)\sim \frac{h}{(z-z_1)^2}+\frac{\gamma_1}{z-z_1}+..., \notag \\
&~& T_1(z)\sim \frac{h}{(z-z_2)^2}+\frac{\gamma_2}{z-z_2}+... .\eea
Then the differential equation on $v_1(z)$ is of the form
\be \frac{\partial}{\partial z}v_1(z)=-v_0(z)^{-1}a_1(z)v_0(z)v_1(z), \ee
which could be  solved formally as a path-ordered integral
\be v_1(z)={\cal{P}}\exp\left(-\int dz v_0(z)^{-1}a_1(z) v_0(z) \right). \ee
To the leading order of $h$ the trivial monodromy condition requires that
\be \oint dz v_0(z)^{-1} a_1(z) v_0(z)=0. \ee
By the residue theorem, it leads to the following monodromy equation
\be\label{diff} {\cal M}=h\frac{\partial}{\partial z}P(z)\mid_{z_1}+\gamma_1P(z_1)
+h\frac{\partial}{\partial z}P(z)\mid_{z_2} +\gamma_2 P(z_2)=0, \ee
where
\be P(z)=v_0(z)^{-1}\left(
\begin{array}{ccc}
0& 1 \\
0 &0
\end{array} \right) v_0(z),
\ee
and
\be \frac{\partial }{\partial z}P(z)=v_0(z)^{-1}
\left(a_0(z)
\left( \begin{array}{ccc}
0 &1 \\
0& 0
\end{array} \right)
-\left( \begin{array}{ccc}
0 &1 \\
0& 0
\end{array} \right)
a_0(z) \right) v_0(z)
=v_0(z)^{-1} \left(
\begin{array}{ccc}
-1&0 \\
0& 1
\end{array} \right)
v_0(z).
\ee
 In principle, one can solve the parameters $\gamma_1$ and $\gamma_2$ from the equation (\ref{diff}). Then the two-point function can be read from the following differential equations
\bea\label{gamma} &~&\frac{\partial }{\partial z_1} \langle \phi_1(z_1) \phi_2(z_2)\rangle_b=\gamma_1,
\notag \\
&~& \frac{\partial }{\partial z_2} \langle \phi_1(z_1) \phi_2(z_2)\rangle_b=\gamma_2. \eea
In practice, it is not easy to find $\g_1,\g_2$. Instead of directly solving the equation (\ref{diff}), we will use the holographic method to compute the two-point function and read the parameters $\g_1,\g_2$, and then check that it  satisfies the monodromy equation.

\section{Holographic computation via  Wilson line}

From the AdS$_3$/CFT$_2$ correspondence, the large $c$ limit of the 2D CFT corresponds to the semiclassical AdS$_3$ gravity. The leading order partition function on the boundary Riemann surface can be read from the on-shell regularized action of the gravitational configuration ending on the Riemann surface.
This picture leads to the proof of the RT formula for the multi-interval entanglement entropy in the vacuum state of the CFT. However for a general state in CFT, this approach may  not be effective.

On the other hand, it has been know for a long time that the pure AdS$_3$ gravity could be cast into a Chern-Simons form\cite{Achucarro:1987vz,Witten:1988hc}.
And it was proposed \cite{Ammon:2013hba} that the holographic entanglement entropy could be read from the probe action of the Wilson line. In this section, we follow the Wilson-line prescription to compute the HEE in the large class of CFT states.
%In this section, we will briefly review the Wilson line prescription for holographic entanglement entropy and the relation with geodesic for pure gravity following . We will first focus on pure gravity case and then extend it to higher spin gravity.

\subsection{Wilson line prescription}

It is well known that the three dimensional AdS$_3$ gravity has no local degree of freedom and can be rewritten into a  Chern-Simons theory\cite{Achucarro:1987vz,Witten:1988hc}
\be S=S_{CS}[A]-S_{CS}[\bar{A}], \ee
where
\be S_{CS}[A]=\frac{k}{4\pi}\int \Tr(A\wedge dA+A\wedge A\wedge A), \ee
and both  gauge potentials takes valued in the $sl(2)$ algebra. The metric can be read out by
\be g_{\mu \nu}=\frac{1}{2} \Tr (A-\bar{A})_{\mu}(A-\bar{A})_{\nu}. \ee
The gauge symmetry in the Chern-Simons theory is related to the local Lorentz symmetry and diffeomorphism invariance of the gravity.
%By changing the gauge transformation to $SL(N)\times SL(N)$, the theory can be extended to higher spin theory.

It is suggested that the holographic entanglement entropy can be evaluated as the action of a Wilson line probe in the Chern-Simons  theory \cite{Ammon:2013hba}
\be S_{HEE}=-\log (W_{\cal{R}}({\cal{C}})), \label{WEE}\ee
where ${\cal{C}}$ is the bulk Wilson line ending at the branch points of the interval. The representation for gauge potential in the Wilson line is of infinite dimension,   whose Casimir is related to the conformal dimension of the twist operator. The infinite dimensional representation can be generated by an auxiliary field on the Wilson line\cite{Witten:1989sx}
\be S(U,P)=\int \Tr PU^{-1}\frac{d}{ds}U+\lambda(s)(\Tr (P^2)-c_2), \ee
where $P$, $U$, $\lambda$ are the auxiliary fields  and $c_2$ is  the Casimir
\be c_2=2h(h-1), \ee
where $h$ is the $L_0$ eigenvalue of the highest weight state.
The auxiliary field theory can be coupled to the gauge field as
\be\label{action} S(U,P,A,\bar{A})=\int \Tr PU^{-1}D_sU+\lambda(s)(\Tr (P^2)-c_2), \ee
where
\be D_s U=\frac{d}{ds} U+A_s U-U\bar{A}_s,~~~A_s=A_{\mu}\frac{dx^{\mu}}{ds}. \ee
The Wilson line for the highest weight representation can be obtained by taking a path integral over the auxiliary field theory
\bea\label{path} &~&W_{\cal{R}}({\cal{C}})=\tr_{{\cal{R}}}\left( {\cal{P}}\exp (\oint_{{\cal{C}}} A)\right)=
{\int} [dU dP] e^{-S(U,P,A)}, \notag \\
&~&W_{\cal{R}}({\cal{C}}_{ij})=\langle j\mid {\cal{P}}\exp (\int_{{\cal{C}}_{ij}} A)\mid i\rangle=
{\int} [dU dP] e^{-S(U,P,A)_{ij}}. \eea
However there are some subtleties in using the second equation of (\ref{path}) to study the entanglement entropy. First of all, it is not clear what are the bra state and the ket state inserting at the end points. Secondly the relation between the states inserting at end points and the boundary condition for the auxiliary fields are not clear. In \cite{Ammon:2013hba}, the authors suggested that the boundary condition should be
\be U_i=U_f=1. \ee
%{\it With this boundary condition, they has  shown that for the Chern-Simons theory with $SL(2)\times SL(2)$ gauge group,  the  entropy computed by (\ref{WEE}) reproduces  the Ryu-Takayanagi formula.}

In the semi-classical limit the path integral (\ref{path}) can be evaluated by the saddle point approximation. Taking the variation with respect to $P$ and $U$, we have  the equations of motion
\bea\label{equ} &~&U^{-1}D_sU+2\lambda P=0, \notag \\
&~& \frac{d}{ds}P+[\bar{A}_s,P]=0, \notag \\
&~& \Tr P^2=c_2. \eea
The on-shell action equals to
\be\label{onshell} S_{\mbox{on-shell}}=\int_{\cal{C}} ds \Tr (PU^{-1}D_s U)=-2c_2\int _{\cal{C}} ds \lambda(s)
=\sqrt{c_2}\int_{\cal{C}}ds \sqrt{\Tr (U^{-1}D_sU)^2}. \ee
Taking a variation of (\ref{onshell}0 with respect to $U$, we get the differential equation
\be\label{equation} \frac{d}{dx}\left(\frac{1}{\sqrt{\Tr(U^{-1}D_sU)^2}}(A_s^{u}-\bar{A}_s)\right)+
\frac{1}{\sqrt{\Tr(U^{-1}D_sU)^2}}[\bar{A}_s,A_s^{u}]=0, \ee
where
\bea &~&A_s^{u}=U^{-1}\frac{d}{ds}U+U^{-1}A_{\mu}U\frac{dx^{\mu}}{ds}, \notag \\
&~& \bar{A}_s=\bar{A}_{\mu}\frac{dx^{\mu}}{ds}. \eea
For the pure gravity case,  the dimension of space-time is the same as the dimension of the $SL(2)$ gauge group. It is possible to choose a special curve such that $U\equiv 1$ on the curve. We parameterize the curve such that $\sqrt{\Tr(U^{-1}D_sU)^2}\equiv 1$. The on-shell action (\ref{onshell}) equals to the length of the curve and the equation of motion (\ref{equation}) reduces to
\be
\frac{d}{ds}\left((A-\bar A)_\mu\frac{dx^\m}{ds}\right)+[\bar A_\m, A_\n]\frac{dx^\m}{ds}\frac{dx^\n}{ds}=0,
\ee
which could be rewritten in terms of the vielbein and the spin connection,
\be \frac{d}{ds}({e_{\mu}}^a \frac{d x^{\mu}}{ds})+{{\omega_{\mu}}^a}_b {e_{\nu}}^b
\frac{d x^{\mu}}{ds}\frac{d x^{\nu}}{ds}=0. \ee
This is exactly the geodesic equation. On the other hand, the on-shell action is given by the  length of the geodesic ending on the branch points. Therefore, the on-shell action of the Wilson-line probe gives the RT formula.

%\subsection{Higher spin}
There are two remarkable points in the above discussion. One point is that the on-shell action is independent of the path connecting two branch points. Different choice of the path are gauge equivalent to each other. The other point is that  the equivalence between the on-shell action and the geodesic length is general, independent of the background spacetime. It is this second point that allows us to  prove the RT formula for a large class of CFT states.

%\section{Monodromy prescription in Virasoro algebra}

\subsection{RT=EE in a large class of CFT states}

Based on the holographic dictionary \cite{Kraus:2006wn}, the gauge potentials corresponding to the boundary stress tensor $T_0(z)$ are of the forms
\bea &~&A=(e^{\rho}L_1+\frac{6}{c}T_0(z)e^{-\rho}L_{-1})dz+L_0 d\rho, \notag \\
&~& \bar{A}=(e^{\rho}L_{-1}+\frac{6}{c}\tilde{T}_0(\bar{z})e^{-\rho}L_1)d\bar{z} -L_0 d\rho. \label{Banados}\eea
Note that we do not require the boundary stress tensor to be constant. Consequently, the gravitational configurations are quite general, the so-called Banados geometry\cite{Banados:1998gg}.
The on-shell action (\ref{onshell}) can be evaluated by the Wilson-line prescription \cite{Ammon:2013hba}. The gauge potential is flat and can be locally related to the ``nothingness" solution $A=0$ by a gauge transformation
\be
A=Vd V^{-1}, \hs{3ex}\bar A=\bar V d\bar V^{-1}. \label{gaugetr}
\ee
Put it in another way, we have
\bea &~&\partial_{\mu}V=-A_{\mu} V , \notag \\
&~& \partial_{\mu}\bar{V}=-\bar{A}_{\mu} \bar{V}, \eea
which have the solutions
\bea &~&V(x_f,x_i)=e^{-\rho_{f} L_0}v_0(z_f,z_i)e^{\rho_i L_0}, \notag \\
&~& \bar{V}(x_f,x_i)=e^{\rho_f L_0}\bar{v}_0(\bar{z}_f,\bar{z}_i)e^{-\rho_i L_0}. \eea
Here we have introduced
\be v_0(z_f,z_i)=v_0(z_f)v_0(z_i)^{-1}, \hs{3ex}\bar v_0(z_f,z_i)=\bar v_0(z_f)\bar v_0(z_i)^{-1},\ee
in which the quantity $v_0(z)$ is the solution of the equation (\ref{v0}) and similarly for $\bar v_0$ in the anti-holomorphic sector. With the gauge potential (\ref{gaugetr}), the solutions to the equation of motion (\ref{equ}) are
\bea &~&U(s)=V(x(s),x_i)U_0(s)\bar{V}(x(s),x_i)^{-1}, \notag \\
&~& P(s)=\bar{V}(x(s),x_i)P_0(s)\bar{V}(x(s),x_i)^{-1},
\eea
where
\bea &~&U_0(s)=U_iexp[-\int_{s_i}^{s} ds 2\lambda(s) P_0], \notag \\
&~&P_0(s)=\mbox{const}. \eea
As shown in \cite{Ammon:2013hba}, the on-shell action (\ref{onshell}) can be evaluated by the initial and final values of $U(s)$. We can define a matrix as
\bea\label{M} M&=&U_0(s_f)^{-1}U_0(s_i) \notag \\
&=&\exp[\int_{s_i}^{s_f}ds 2\lambda(s) P_0] \notag \\
&\sim &U_f^{-1}V(x_f,x_i)^{-1}U_i\bar{V}(x_f,x_i),\eea
which can be diagnosed  $HMH^{-1}=(\lambda,\frac{1}{\lambda})$. The on-shell action equals to
\be S_{\mbox{on-shell}}=\Tr \log(( HMH^{-1})\sqrt{2c_2}L_0). \ee
The eigenvalue $\lambda$ of $M$ is captured by the trace of the matrix
\bea\label{Sonshell} \tr M&\sim& e^{2\rho_{\inf}}\tr
\left(\left( \begin{array}{ccc}
0 & 0 \\
0& 1
\end{array} \right)
v_0(z_f,z_i)
\left( \begin{array}{ccc}
1 &0 \\
0 &0
\end{array} \right)
\bar{v}_0(\bar{z}_f,\bar{z}_i)^{-1}\right) \\
&=& e^{2\rho_{\inf}} \tr \left(v_0(z_f,z_i)
\left(
\begin{array}{ccc}
0 &1 \\
0 &0
\end{array} \right)\right)
\tr \left(\left(
\begin{array}{ccc}
0 &0 \\
1& 0
\end{array}
\right) \bar{v}_0(\bar{z}_f,\bar{z}_i)^{-1}\right),
\eea
where we only keep the leading order of the IR cut-off so that the trace is factorized. Thus the eigenvalue can be approximated by
\be \lambda\sim \tr M, \ee
to the leading order of IR cut-off.

The correlation function of the two light operators  and the entanglement entropy can be read from the on-shell action of  the Wilson line. We just need to set $\sqrt{2c_2}\rightarrow 2h$ for the two-point function, and $\sqrt{2c_2}\rightarrow \frac{c}{6}$ for the entanglement entropy.  It is expected that  the two-point function in CFT can be evaluated holographically by
\bea \lefteqn{\log \langle \phi(z_i,\bar{z}_i) \phi(z_f,\bar{z}_f) \rangle_b
=-2h \tr (\log(HMH^{-1})L_0)} \notag \\
&=&-2h (\log \tr(v_0(z_f,z_i)
\left(
\begin{array}{ccc}
0 &1 \\
0& 0
\end{array}
\right)
+ \log \tr
\left( \begin{array}{ccc}
0 &0 \\
1& 0
\end{array}
\right) \bar{v}_0(\bar{z}_f,\bar{z}_i)^{-1} )+\mbox{const}.
\eea
We now show that this is true. First we work out the accessory parameters by setting $z_f=z_1, z_i=z_2$ and using
 (\ref{gamma}), and we get
\bea &~&\gamma_1=-2h\frac{\tr \frac{\partial}{\partial z_1} v_0(z_1)v_0(z_2)^{-1}
\left(
\begin{array}{ccc}
0 & 1\\
0& 0
\end{array} \right) }
{K_0}=2h\frac{K_1 }
{K_0}, \notag \\
&~& \gamma_2=-2h
\frac{\tr  \frac{\partial}{\partial z_2}v_0(z_1)v_0(z_2)^{-1}
\left( \begin{array}{ccc}
0 &1 \\
0& 0
\end{array} \right)}
{K_0}
=-2h \frac{K_2 }
{K_0 }.
\eea
where
\bea
K_0&=&\tr v_0(z_1)v_0(z_2)^{-1}
\left( \begin{array}{ccc}
0 &1 \\
0 &0
\end{array} \right),\nn\\
K_1&=& \tr  v_0(z_1)v_0(z_2)^{-1}
\left(
\begin{array}{ccc}
1 & 0\\
0& 0
\end{array} \right),\nn\\
K_2&=&\tr v_0(z_1)v_0(z_2)^{-1}
\left(
\begin{array}{ccc}
0 &0 \\
0 &1
\end{array}
\right).
\eea
Then we check that these two parameters do satisfy the monodromy condition.
Taking them into the equation (\ref{diff}), we find the left-hand side of the equation
\bea{\cal M}&=&h v_0(z_1)^{-1}
\left( \begin{array}{ccc}
-1 & 0 \\
0& 1
\end{array} \right)
v_0(z_1)
+2h \frac{K_1 }
{K_0}
v_0(z_1)^{-1}
\left( \begin{array}{ccc}
0& 1 \\
0& 0
\end{array} \right)
v_0(z_1) \notag \\
&~&+h v_0(z_2)^{-1}
\left( \begin{array}{ccc}
-1 & 0 \\
0& 1
\end{array} \right) v_0(z_2)
-2h\frac{ K_2}
{K_0}
v_0(z_2)^{-1}
\left( \begin{array}{ccc}
0 &1 \\
0& 0
\end{array}
\right) v_0(z_2). \notag \\
\eea
To prove ${\cal M}=0$ we only need to take a trace with $v_0(z_2)^{-1}R v_0(z_1)$, where $R$ can be a set of linearly independent two by two matrices. We choose
\be R=\mathbf{I},~
\left(\begin{array}{ccc}
1& 0 \\
0& -1
\end{array}\right),~
\left(\begin{array}{ccc}
0& 1\\
0& 0
\end{array}\right),~
\left( \begin{array}{ccc}
0& 0 \\
1& 0
\end{array} \right). \ee
It is easy to show ${\cal M}=0$. In other words, the accessory parameters determined by the on-shell action of the Wilson line satisfy the trivial monodromy condition.  Consequently we prove that the on-shell action of the Wilson line gives correctly the two-point function of two light operators. Considering the fact that the on-shell action of the Wilson line gives the Ryu-Takayanagi formula, we prove that the RT formula captures the single interval entanglement entropy in a large class of states of CFT.

%\section{Higher spin case}

\section{Conclusion and Discussion}

In this work, we studied the entanglement entropy for a large class states in the 2D large $c$  CFT. The  state could be the state generated by finite number of the local operators, or the thermal state generated by a thermal density matrix, or both. In the large $c$ limit, we still assume that the vacuum conformal block dominates the contribution so that the problem can be studied by the monodromy analysis on a differential equation.  On the bulk side, we showed that the on-shell action of the Wilson-line probe captures the entropy correctly. Due to the equivalence between the on-shell Wilson-line action and the geodesic length, the RT formula in  the background corresponding the general state must reproduce the entanglement entropy to the leading order as well.

Our study can be generalized to the higher spin case straightforwardly. Within the Chern-Simons formulation, it is easy to couple the three dimensional gravity to the higher spin fields. By extending the gauge group to $SL(N)\times SL(N)$, the theory describe the higher spin AdS$_3$ gravity with the spin from $2$ up to $N$. It has been proved that with the generalized Brown-Henneaux boundary condition the asymptotic symmetry group for the higher spin gravity is the ${\cal{W}}_N\times {\cal{W}}_N$ algebra \cite{Campoleoni:2010zq,Henneaux:2010xg}. In the higher spin gravity, the usual geometric notions do not make much sense as the gauge transformation in the theory mix the metric field with the higher spin fields. As a result, the geodesic and its length cannot be well-defined. The RT formula for the holographic entanglement entropy has to be modified. It was proposed in \cite{Ammon:2013hba,deBoer:2013vca} that in the higher spin gravity the holographic entanglement entropy can be computed by the on-shell action of an open Wilson line, just as we reviewed before. The Wilson line prescription for the higher spin gravity has been supported from studying the correlation functions via the conformal block in the field theory, both on complex plane \cite{deBoer:2014sna} and on the torus\cite{Chen:2016uvu}.

 We can generalize the discussion on the entanglement entropy in a large class of states to the theory with a $\cW$ symmetry. For the state generated by the operators, the discussion is similar to the one in the above sections. It turns out that the Wilson-line prescription is still effective to read the entanglement entropy. We are not going to repeat the analysis here.   In \cite{Chen:2016uvu}, we studied the entanglement entropy in the $\cW$ theory with the higher spin deformation. In that case, the density matrix consists of not only a finite temperature, but also a higher spin chemical potential
 \be\label{density} \rho=e^{2\pi i\tau{\cal{L}}_0+2\pi i\alpha{\cal{W}}_0}
e^{-2\pi i\bar{\tau}\bar{\cal{L}}_0-2\pi i\bar{\alpha}\bar{\cal{W}}_0}.\ee
The deformation  theory can be related to the original theory by a picture-changing transformation. As a payback, we have to introduce two extra auxiliary coordinates by
\be\label{new} \phi(z,y;\bar{z},\bar{y})\equiv
e^{-i{\cal{W}}_0 y}e^{i{\bar{\cal{W}}}_0\bar{y}}\phi(z,\bar{z}) e^{i{\cal{W}}_0y}e^{i\bar{\cal{W}}_0\bar{y}}, \ee
where the translations along $y,\bar y$  are related to the evolution respect to the charge ${\cW}_0$. The entanglement entropy can be computed by applying the monodromy techniques based on the ${\cW}_3$ algebra. More interestingly,
 it has been shown that the Wilson-line prescription can be reliably applied to compute the entanglement entropy in the field theory, with the gauge potential including the  chemical potential. In the case with the chemical potential, we may consider the entanglement entropy of a more general state in CFT and show that the Wilson line prescription is effective.  More interestingly, inspired by the case with the chemical potential, we consider  the two-point function of the operator  $e^{yW_{-2}}e^{\bar{y}\bar{W}_{-2}}\phi(z,\bar{z})$, which is a special kind of descendant operator.
We find that the two-point function can still be computed in the  Wilson-line prescription, but now the Wilson line is defined in five dimensions, including two extra dimensions.

%We admit that, we don't fully understand the physics meaning of the five dimensional space. Up to now, it is only an effective way to calculate the two descendant operators' correlation function holographically. On the other hand, we suggest there are other way to calculate the correlation function in three dimensional space. Recall the Wilson line prescription \ref{matrix} and \ref{twopoint}, there are still freedom for $U_i$ and $U_f$. In this case we just set $U_i=U_f=1$. For descendant operator, it is possible to have a different initial and final condition for $U$ field.

%In ${\cal{W}}_3$ algebra, there are more complicate descendant operators. The correlation function with those operator is still not clear. We conjecture that the operator like $e^{y^{(1)}W_{-1}}e^{y^{(2)}W_{-2}}\phi(z)$ may have simple effect. In large $c$ limit, the operator $L_{\pm 1}~L_0~W_{\pm 2}~W_{\pm 1}~W_0$ become a closed sub-algebra. It is expected that the states generated by the sub-algebra may have some simple effect. We leave the discussion to future.

Our discussion is quite general and can be applied to  other cases. The only assumption is that the vacuum module states dominates the contribution in the operator product expansion of two operators. This sets the trivial monodromy condition for the cycle enclosing these two operators. As a result, the classical order part of the two-point function can be calculated, based on  the expectation values of the stress tensor and $W$ current.  On the bulk side, the expectation values of the stress tensor and the $W$ current appears in the background gauge potential, which is essential to compute the on-shell action of the Wilson-line probe.  Different from the cases discussed in the literature \cite{Ryu:2006bv}, the expectation values can be non-constant such that the dual background spacetime could be a general Banados spacetime\cite{Banados:1998gg} and its higher spin generalization.   It is interesting to apply this prescription to study the multi-shockwave effect and the shockwave in the higher spin black hole, which are hard to calculate directly.

Our result is also related to the work studying the linear gravitional equation from the entanglement entropy \cite{Blanco:2013joa}. The second order result has been studied recently in \cite{Beach:2016ocq}. Our result is an support for their study in the three dimensional gravity.

One essential assumption under our study is that the gravitational dual of the CFT state is the Banados spacetime (\ref{Banados})\footnote{We would like to thank the anonymous referee for the comments on the CFT states, which help us to improve our statement.}. Consequently our discussion does not apply for a general state whose dual is not clear or not of the form as (\ref{Banados}). One notorious case is that for two CFT state $|\psi_1\rangle$ and $|\psi_2\rangle$, each of which has a  dual description  (\ref{Banados}) with $T_1(z)$ and $T_2(z)$ respectively, but it is not clear in general what is the gravitational dual to the superposed state $\a_1|\psi_1\rangle+\a_2|\psi_2\rangle$. It is certain that the dual configuration cannot be the one with $T_1(z)+T_2(z)$. Another case is for the collapse state for the black hole formation, studied in \cite{Anous:2016kss}. In this case, the collapse state is a continuous distribution of the operators on a circle, and its dual is the AdS$_3$-Vaidya geometry. The collapse state does not belong to the class of the states in  our treatment, though the RT formula still holds in this case.  It would be interesting to study the RT formula for a more general state in CFT.

%\newpage

\vspace*{10mm}
\noindent {\large{\bf Acknowledgments}}\\

The work was in part supported by NSFC Grant No.~11275010, No.~11335012 and No.~11325522.
We would like to thank T. Takayanagi and J. Long for helpful discussions. J.Q. Wu was supported by Short-term Overseas Research Program Graduate School of Peking University. We would like to thank the anonymous referee for the valuable suggestions and comments. 
%%The work was in part supported by NSFC Grant No. 10975005.
\vspace*{5mm}

\begin{appendix}

\end{appendix}


\vspace*{5mm}
\begin{thebibliography}{99}

\bibitem{nielsen2010quantum}
M.~A. Nielsen and I.~L. Chuang, {\em Quantum computation and quantum
  information}.
\newblock Cambridge university press, 2010.

\bibitem{petz2008quantum}
D.~Petz, {\em Quantum information theory and quantum statistics}.
\newblock Springer, 2008.

\bibitem{Ryu:2006bv}
  S.~Ryu and T.~Takayanagi,
  ``Holographic derivation of entanglement entropy from AdS/CFT,''
  Phys.\ Rev.\ Lett.\  {\bf 96}, 181602 (2006)
  doi:10.1103/PhysRevLett.96.181602
  [hep-th/0603001].

  \bibitem{Ryu:2006ef}
  S.~Ryu and T.~Takayanagi,
  ``Aspects of Holographic Entanglement Entropy,''
  JHEP {\bf 0608}, 045 (2006)
  doi:10.1088/1126-6708/2006/08/045
  [hep-th/0605073].

\bibitem{Casini:2011kv}
  H.~Casini, M.~Huerta and R.~C.~Myers,
  ``Towards a derivation of holographic entanglement entropy,''
  JHEP {\bf 1105}, 036 (2011)
  doi:10.1007/JHEP05(2011)036
  [arXiv:1102.0440 [hep-th]].

  \bibitem{Lewkowycz:2013nqa}
  A.~Lewkowycz and J.~Maldacena,
  ``Generalized gravitational entropy,''
  JHEP {\bf 1308}, 090 (2013)
  doi:10.1007/JHEP08(2013)090
  [arXiv:1304.4926 [hep-th]].

\bibitem{Headrick:2010zt}
  M.~Headrick,
  ``Entanglement Renyi entropies in holographic theories,''
  Phys.\ Rev.\ D {\bf 82}, 126010 (2010)
  [arXiv:1006.0047 [hep-th]].

  \bibitem{Faulkner:2013yia}
  T.~Faulkner,
  ``The Entanglement Renyi Entropies of Disjoint Intervals in AdS/CFT,''
  arXiv:1303.7221 [hep-th].
  %%CITATION = ARXIV:1303.7221;%%

\bibitem{Hartman:2013mia}
  T.~Hartman,
  ``Entanglement Entropy at Large Central Charge,''
  arXiv:1303.6955 [hep-th].




\bibitem{Faulkner:2013ana}
  T.~Faulkner, A.~Lewkowycz and J.~Maldacena,
  ``Quantum corrections to holographic entanglement entropy,''
  JHEP {\bf 1311}, 074 (2013)
  doi:10.1007/JHEP11(2013)074
  [arXiv:1307.2892 [hep-th]].

  \bibitem{Barrella:2013wja}
  T.~Barrella, X.~Dong, S.~A.~Hartnoll and V.~L.~Martin,
  ``Holographic entanglement beyond classical gravity,''
  JHEP {\bf 1309}, 109 (2013)
  doi:10.1007/JHEP09(2013)109
  [arXiv:1306.4682 [hep-th]].

\bibitem{Krasnov:2000zq}
  K.~Krasnov,
  ``Holography and Riemann surfaces,''
  Adv.\ Theor.\ Math.\ Phys.\  {\bf 4}, 929 (2000)
  [hep-th/0005106].

\bibitem{Yin:2007gv}
  X.~Yin,
  ``Partition Functions of Three-Dimensional Pure Gravity,''
  Commun.\ Num.\ Theor.\ Phys.\  {\bf 2}, 285 (2008)
  [arXiv:0710.2129 [hep-th]].

  \bibitem{Giombi:2008vd}
  S.~Giombi, A.~Maloney and X.~Yin,
  ``One-loop Partition Functions of 3D Gravity,''
  JHEP {\bf 0808}, 007 (2008)
  [arXiv:0804.1773 [hep-th]].

   \bibitem{Chen:2015uga}
  B.~Chen and J.~q.~Wu,
  ``1-loop partition function in AdS$_{3}$/CFT$_{2}$,''
  JHEP {\bf 1512}, 109 (2015)
  doi:10.1007/JHEP12(2015)109
  [arXiv:1509.02062 [hep-th]].

\bibitem{Holzhey:1994we}
  C.~Holzhey, F.~Larsen and F.~Wilczek,
  ``Geometric and renormalized entropy in conformal field theory,''
  Nucl.\ Phys.\ B {\bf 424}, 443 (1994)
  doi:10.1016/0550-3213(94)90402-2
  [hep-th/9403108].




\bibitem{Zamolodchikov1}
A.~B.~Zamolodchikov,
``Two-dimensional conformal symmetry and critical four-spin correlation functions in the Ashkin-Teller model,'' Sov.\ Phys.\ JETP {\bf 63}, (1986) 1061¨C1066.

\bibitem{Zamolodchikov2}
A.~B.~Zamolodchikov,
``Conformal symmetry in two-dimensional space: Recursion representation of conformal block,''
Theoretical and Mathematical Physics 73 no. 1, 1088¨C1093.

\bibitem{Fitzpatrick:2014vua}
  A.~L.~Fitzpatrick, J.~Kaplan and M.~T.~Walters,
  ``Universality of Long-Distance AdS Physics from the CFT Bootstrap,''
  JHEP {\bf 1408}, 145 (2014)
  doi:10.1007/JHEP08(2014)145
  [arXiv:1403.6829 [hep-th]].

  \bibitem{Asplund:2014coa}
  C.~T.~Asplund, A.~Bernamonti, F.~Galli and T.~Hartman,
  ``Holographic Entanglement Entropy from 2d CFT: Heavy States and Local Quenches,''
  JHEP {\bf 1502}, 171 (2015)
  doi:10.1007/JHEP02(2015)171
  [arXiv:1410.1392 [hep-th]].
  
 \bibitem{Nozaki:2014hna} 
  M.~Nozaki, T.~Numasawa and T.~Takayanagi,
  ``Quantum Entanglement of Local Operators in Conformal Field Theories,''
  Phys.\ Rev.\ Lett.\  {\bf 112}, 111602 (2014)
  doi:10.1103/PhysRevLett.112.111602
  [arXiv:1401.0539 [hep-th]]. M.~Nozaki,
  ``Notes on Quantum Entanglement of Local Operators,''
  JHEP {\bf 1410}, 147 (2014)
  doi:10.1007/JHEP10(2014)147
  [arXiv:1405.5875 [hep-th]]. S.~He, T.~Numasawa, T.~Takayanagi and K.~Watanabe,
  ``Quantum dimension as entanglement entropy in two dimensional conformal field theories,''
  Phys.\ Rev.\ D {\bf 90}, no. 4, 041701 (2014)
  doi:10.1103/PhysRevD.90.041701
  [arXiv:1403.0702 [hep-th]].B.~Chen, W.~Z.~Guo, S.~He and J.~q.~Wu,
  ``Entanglement Entropy for Descendent Local Operators in 2D CFTs,''
  JHEP {\bf 1510}, 173 (2015)
  doi:10.1007/JHEP10(2015)173
  [arXiv:1507.01157 [hep-th]].

\bibitem{Caputa:2014vaa} 
  P.~Caputa, M.~Nozaki and T.~Takayanagi,
  ``Entanglement of local operators in large-N conformal field theories,''
  PTEP {\bf 2014}, 093B06 (2014)
  doi:10.1093/ptep/ptu122
  [arXiv:1405.5946 [hep-th]].
  
 \bibitem{Caputa:2014eta} 
  P.~Caputa, J.~Simón, A.~¦tikonas and T.~Takayanagi,
  ``Quantum Entanglement of Localized Excited States at Finite Temperature,''
  JHEP {\bf 1501}, 102 (2015)
  doi:10.1007/JHEP01(2015)102
  [arXiv:1410.2287 [hep-th]].

  \bibitem{Caputa:2015waa}
  P.~Caputa, J.~Simón, A.~¦tikonas, T.~Takayanagi, and K.~Watanabe,
  ``Scrambling time from local perturbations of the eternal BTZ black hole,''
  JHEP {\bf 1508}, 011 (2015)
  doi:10.1007/JHEP08(2015)011
  [arXiv:1503.08161 [hep-th]].




  \bibitem{Roberts:2014ifa}
  D.~A.~Roberts and D.~Stanford,
  ``Two-dimensional conformal field theory and the butterfly effect,''
  Phys.\ Rev.\ Lett.\  {\bf 115}, no. 13, 131603 (2015)
  doi:10.1103/PhysRevLett.115.131603
  [arXiv:1412.5123 [hep-th]].

  \bibitem{Fitzpatrick:2016thx}
  A.~L.~Fitzpatrick and J.~Kaplan,
  ``A Quantum Correction To Chaos,''
  arXiv:1601.06164 [hep-th].

  \bibitem{Chen:2016uvu}
  B.~Chen and J.~q.~Wu,
  ``Higher spin entanglement entropy at finite temperature with chemical potential,''
  arXiv:1604.03644 [hep-th].

  \bibitem{deBoer:2014sna}
  J.~de Boer, A.~Castro, E.~Hijano, J.~I.~Jottar and P.~Kraus,
  ``Higher spin entanglement and $ {\mathcal{W}}_{\mathrm{N}} $ conformal blocks,''
  JHEP {\bf 1507}, 168 (2015)
  doi:10.1007/JHEP07(2015)168
  [arXiv:1412.7520 [hep-th]].

  \bibitem{Anous:2016kss}
  T.~Anous, T.~Hartman, A.~Rovai and J.~Sonner,
  ``Black Hole Collapse in the 1/c Expansion,''
  arXiv:1603.04856 [hep-th].

 \bibitem{Chen:2014unl}
  B.~Chen and J.~q.~Wu,
  ``Single interval Renyi entropy at low temperature,''
  JHEP {\bf 1408}, 032 (2014)
  doi:10.1007/JHEP08(2014)032
  [arXiv:1405.6254 [hep-th]].

  \bibitem{Chen:2015kua}
  B.~Chen and J.~q.~Wu,
  ``Holographic calculation for large interval R¨¦nyi entropy at high temperature,''
  Phys.\ Rev.\ D {\bf 92}, no. 10, 106001 (2015)
  doi:10.1103/PhysRevD.92.106001
  [arXiv:1506.03206 [hep-th]].

   \bibitem{Chen:2015uia}
  B.~Chen, J.~q.~Wu and Z.~c.~Zheng,
  ``Holographic R\'enyi Entropy of Single Interval on Torus: with W symmetry,''
  Phys.\ Rev.\ D {\bf 92}, 066002 (2015)
  [arXiv:1507.00183 [hep-th]].

 \bibitem{Azeyanagi:2007bj}
  T.~Azeyanagi, T.~Nishioka and T.~Takayanagi,
  ``Near Extremal Black Hole Entropy as Entanglement Entropy via AdS(2)/CFT(1),''
  Phys.\ Rev.\ D {\bf 77}, 064005 (2008)
  doi:10.1103/PhysRevD.77.064005
  [arXiv:0710.2956 [hep-th]].

\bibitem{Sheikh-Jabbari:2016znt} 
  M.~M.~Sheikh-Jabbari and H.~Yavartanoo,
  ``Excitation Entanglement Entropy in 2d Conformal Field Theories,''
  arXiv:1605.00341 [hep-th].
  
  \bibitem{Ammon:2013hba}
  M.~Ammon, A.~Castro and N.~Iqbal,
  ``Wilson Lines and Entanglement Entropy in Higher Spin Gravity,''
  JHEP {\bf 1310}, 110 (2013)
  doi:10.1007/JHEP10(2013)110
  [arXiv:1306.4338 [hep-th]].
 
   
   \bibitem{Achucarro:1987vz}
  A.~Achucarro and P.~K.~Townsend,
  ``A Chern-Simons Action for Three-Dimensional anti-De Sitter Supergravity Theories,''
  Phys.\ Lett.\ B {\bf 180}, 89 (1986).
  doi:10.1016/0370-2693(86)90140-1

  \bibitem{Witten:1988hc}
  E.~Witten,
  ``(2+1)-Dimensional Gravity as an Exactly Soluble System,''
  Nucl.\ Phys.\ B {\bf 311}, 46 (1988).
  doi:10.1016/0550-3213(88)90143-5

 
 \bibitem{Witten:1989sx}
  E.~Witten,
  ``Topology Changing Amplitudes in (2+1)-Dimensional Gravity,''
  Nucl.\ Phys.\ B {\bf 323}, 113 (1989).
  doi:10.1016/0550-3213(89)90591-9

    \bibitem{Kraus:2006wn}
  P.~Kraus,
  ``Lectures on black holes and the AdS(3)/CFT(2) correspondence,''
  Lect.\ Notes Phys.\  {\bf 755}, 193 (2008)
  [hep-th/0609074].

\bibitem{Banados:1998gg}
  M.~Banados,
  ``Three-dimensional quantum geometry and black holes,''
  AIP Conf.\ Proc.\  {\bf 484}, 147 (1999)
  doi:10.1063/1.59661
  [hep-th/9901148].


  \bibitem{Campoleoni:2010zq}
  A.~Campoleoni, S.~Fredenhagen, S.~Pfenninger and S.~Theisen,
  ``Asymptotic symmetries of three-dimensional gravity coupled to higher-spin fields,''
  JHEP {\bf 1011}, 007 (2010)
  doi:10.1007/JHEP11(2010)007
  [arXiv:1008.4744 [hep-th]].

  \bibitem{Henneaux:2010xg}
  M.~Henneaux and S.~J.~Rey,
  ``Nonlinear $W_{\infty}$ as Asymptotic Symmetry of Three-Dimensional Higher Spin Anti-de Sitter Gravity,''
  JHEP {\bf 1012}, 007 (2010)
  doi:10.1007/JHEP12(2010)007
  [arXiv:1008.4579 [hep-th]].

  \bibitem{deBoer:2013vca}
  J.~de Boer and J.~I.~Jottar,
  ``Entanglement Entropy and Higher Spin Holography in AdS$_3$,''
  JHEP {\bf 1404}, 089 (2014)
  doi:10.1007/JHEP04(2014)089
  [arXiv:1306.4347 [hep-th]].








  \bibitem{Blanco:2013joa}
  D.~D.~Blanco, H.~Casini, L.~Y.~Hung and R.~C.~Myers,
  ``Relative Entropy and Holography,''
  JHEP {\bf 1308}, 060 (2013)
  doi:10.1007/JHEP08(2013)060
  [arXiv:1305.3182 [hep-th]].

  \bibitem{Beach:2016ocq}
  M.~J.~S.~Beach, J.~Lee, C.~Rabideau and M.~Van Raamsdonk,
  ``Entanglement entropy from one-point functions in holographic states,''
  arXiv:1604.05308 [hep-th].











\end{thebibliography}
\end{document}